\documentclass[runningheads]{llncs}

\usepackage[T1]{fontenc}
\usepackage{graphicx}
\usepackage{amsmath,amssymb}
\usepackage{booktabs}
\usepackage{subcaption}

\begin{document}

\title{Evaluating the Effects of Inter-Observer and Model Variability on Radiological Peritoneal Cancer Index Assessment}
\titlerunning{Inter-Observer and Model Variability in rPCI}

\author{Savvas Saragiotis\inst{1} \and
Pieter C. Gort\inst{1} \and
Lotte J.S. Fleurkens-Ewals\inst{1,2} \and
Anna F. van Herwijnen\inst{1,2} \and
Marion Tops-Welten\inst{2} \and
L.D. Kampmeijer\inst{1,2} \and
Joost Nederend\inst{2} \and
Fons van der Sommen\inst{1}}
\authorrunning{S. Saragiotis et al.}
\institute{Department of Electrical Engineering, Eindhoven University of Technology, Eindhoven, The Netherlands
\email{s.saragiotis@student.tue.nl}\\
\email{\{p.c.gort, fvdsommen\}@tue.nl} \and
Catharina Hospital Eindhoven, Eindhoven, The Netherlands
\email{\{lotte.ewals, marion.tops, lenah.kampmeijer, anna.v.herwijnen, joost.nederend\}@catharinaziekenhuis.nl}}

\maketitle

\begin{abstract}
Deep learning segmentation models are often evaluated using geometric metrics such as Dice, HD95, and ASD, yet it remains unclear to what extent improvements in these metrics translate into clinically meaningful changes in downstream decision-making. The metric-to-decision gap is examined using radiological Peritoneal Cancer Index (rPCI) region segmentation on contrast-enhanced CT, where a consensus definition provides anatomically grounded 3D regions and the clinically used PCI~20 threshold enables decision-level evaluation. Inter-observer variability is quantified across four experts on ten abdominal CT scans, and a published nnU-Net based rPCI segmentation model is benchmarked against this human reference using Dice, HD95, and ASD across all 13 regions. To relate geometric differences to clinical impact, a probabilistic peritoneal metastasis simulation is implemented on majority-vote rPCI maps, propagating region-boundary variability into variability of derived (r)PCI scores and classification at the PCI~20 cutoff. Observers showed high agreement (mean Dice $0.87$), while the model matched human performance in most regions but deviated more in regions 4, 8, and the small-bowel regions (9--12). Across simulations, score differences were typically small (mean $\Delta$rPCI $\approx 0.3$--$0.6$) for both observers and the model, and decision flips occurred predominantly when the reference score was near 20. These results suggest that rPCI-derived scoring is generally robust to typical segmentation variability, while highlighting borderline cases as the main setting where expert review remains essential. 


\end{abstract}

\section{Introduction}
\label{sec:intro}
Deep learning has become the dominant approach for medical image segmentation, and modern architectures achieve strong performance across many tasks ~\cite{Wang2022MedicalImageSegmentationSurvey}. Performance of deep learning models is often measured using geometric similarity metrics such as the Dice Similarity Coefficient (Dice)~\cite{Eelbode2020OptimizationMedicalImageSegmentation}. Although these metrics can quantify the spatial agreement between segmentation masks, they do not directly capture whether a given improvement translates into a clinically meaningful benefit. It is often unclear whether an increase in Dice will correspond to improvements in downstream clinical decision-making, or whether performance differences that appear substantial geometrically have a negligible impact on decisions derived from the segmentation.

Peritoneal metastases (PM) provide a well-suited case study to investigate this metric-to-decision gap. PM are an advanced stage of abdominal cancer in which tumor cells disseminate throughout the peritoneal cavity and implant on peritoneal surfaces~\cite{Rijken2023,Guchelaar2023}. PM commonly arise from colorectal, gastric, ovarian, or pancreatic cancers and are typically associated with poor prognosis~\cite{Spiliotis2015,Benizri2012}. Treatment options such as cytoreductive surgery (CRS) with hyperthermic intraperitoneal chemotherapy (HIPEC) are suitable only for selected patients~\cite{Noiret2022UpdateReview,Guchelaar2023}, making a reliable pre-operative assessment of the disease extent essential.

\begin{figure}[t]
    \centering
    \includegraphics[width=0.9\columnwidth,trim=18mm 70mm 25mm 60mm,clip]{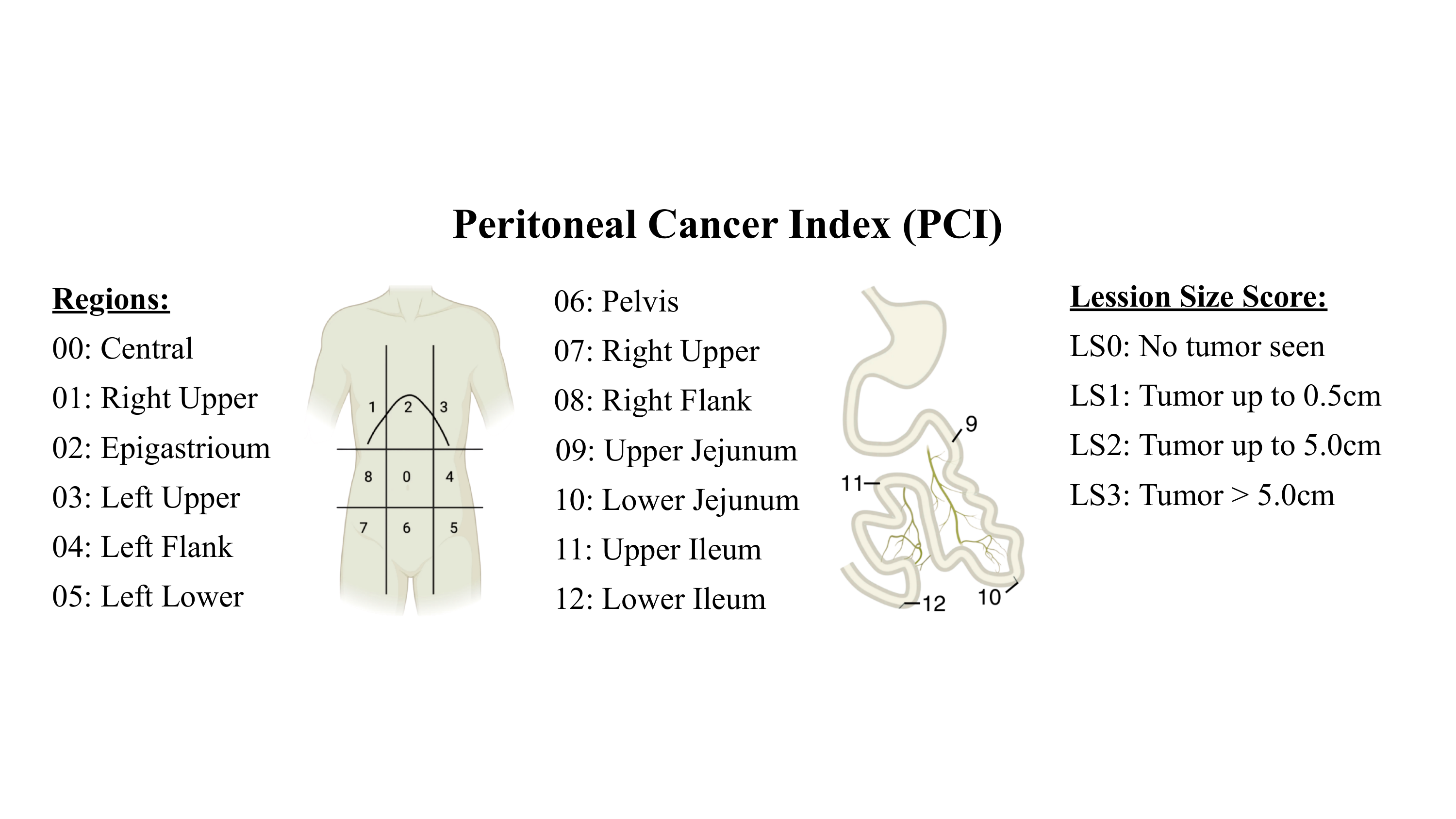}
    \caption{Peritoneal Cancer Index (PCI): anatomical regions and lesion size scoring.}
    \label{fig:PCI}
\end{figure}

The Peritoneal Cancer Index (PCI) is the standard intra-operative scoring system for quantifying PM burden. It divides the abdomen into 13 regions, each scored~0--3 by the largest visible lesion, yielding a total of~0--39 (Fig.~\ref{fig:PCI})~\cite{Jacquet1996ClinicalCarcinomatosis}. The PCI score is used to determine whether a patient is eligible for a certain treatment; for colorectal PM, a PCI of 20 is the most commonly used cutoff for eligibility decisions~\cite{Birgisson2020HighPCI,Noiret2022UpdateReview}. Importantly, this decision boundary enables studying whether segmentation differences can propagate to clinically relevant changes in derived PCI scores and potentially alter treatment classification.

A recent consensus study by Tops-Welten~\emph{et~al.}~\cite{TopsWelten2025} has defined radiological PCI (rPCI) regions on contrast-enhanced CT to facilitate non-invasive and pre-operative estimation of tumor burden. However, rPCI delineation remains challenging, particularly for small-bowel regions where practical boundaries are required~\cite{TopsWelten2025}. Quantifying inter-observer variability is essential to establish a reference range for benchmarking automated approaches and to contextualize model performance beyond geometric metrics~\cite{VeigaCanuto2022,Moliere2024}.

Building on these rPCI definitions, Gort~\emph{et~al.}~\cite{Gort2026rPCISegmentation} trained an nnU-Net model to automatically segment the 13 rPCI regions from contrast-enhanced CT. Although model performance can be summarized by Dice, HD95, and ASD, these metrics alone do not answer the clinically relevant question of whether region-boundary differences meaningfully change derived PCI scores or alter decisions at a treatment threshold. This leads to the key research question: how does model-induced variability in rPCI segmentation compare to human inter-observer variability, and what does this imply for decision-making at a clinically used PCI cutoff?

To address this question, inter-observer variability in rPCI segmentation is quantified on ten abdominal CT scans annotated by four radiologists and compared with the variability introduced by the model proposed in~\cite{Gort2026rPCISegmentation} using Dice, HD95, and ASD. A probabilistic PM simulation is then used to propagate segmentation variability into PCI-score variability and to evaluate its impact on classification around the PCI~20 threshold.

\section{Methods}
\label{sec:meth}
\noindent\textbf{Dataset, Annotations \& Segmentation Model (nnU-Net):}
Ten contrast-enhanced abdominal CT scans were selected from a patient cohort at Catharina Hospital Eindhoven, The Netherlands, including cases with and without confirmed PM. The scans were divided into two cohorts (A and B) of five CT scans each. For each cohort, three clinical researchers independently segmented the 13 rPCI regions using the definitions from the consensus study~\cite{TopsWelten2025}. Four researchers participated in total: two researchers annotated both cohorts, while the other two annotated one cohort each. This resulted in three independent annotations per scan from different annotators across the two cohorts. The segmentation masks were stored as 3D label maps and resampled to $3\,\mathrm{mm}$ isotropic spacing. Furthermore, for all CT scans prediction masks were generated using the model proposed by Gort~\emph{et~al.}~\cite{Gort2026rPCISegmentation}, which was trained for automatic rPCI region segmentation from contrast-enhanced CT. The model was used as provided in that work for inference only; no additional training or fine-tuning was performed in this study. Inference followed the standard nnU-Net pipeline~\cite{Isensee2018} with default post-processing to enable a reproducible comparison against the human segmentations.

\noindent\textbf{Evaluation Metrics:}
Inter-observer and model--observer variability were quantified using the Dice Similarity Coefficient (Dice), which measures volumetric overlap between two segmentations; the 95th percentile Hausdorff distance (HD95), which captures the worst-case boundary deviation while being robust to isolated outliers; and the Average Surface Distance (ASD), which expresses the mean boundary offset in millimeters. Together, Dice reflects global overlap, HD95 highlights local extreme errors, and ASD summarizes typical boundary agreement. 

\begin{figure}[t]
\centering

\begin{subfigure}[b]{0.32\columnwidth}
    \centering
    \includegraphics[width=\linewidth]{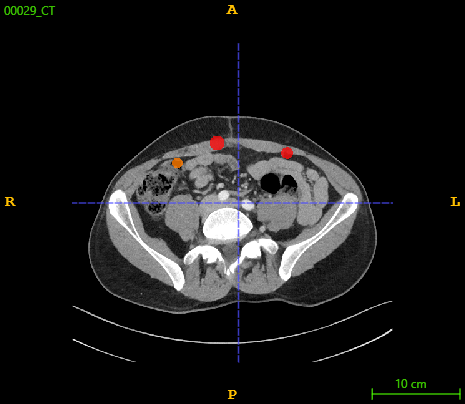}
    \caption{Axial view.}
    \label{fig:sim_pm_axial}
\end{subfigure}\hfill
\begin{subfigure}[b]{0.32\columnwidth}
    \centering
    \includegraphics[width=\linewidth]{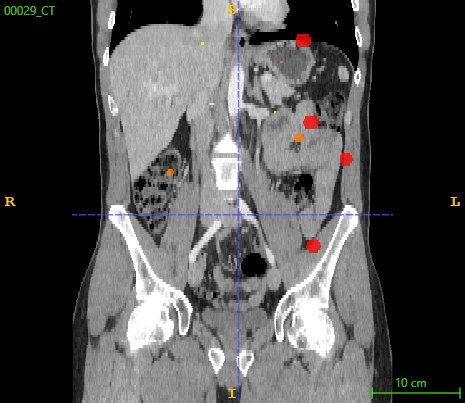}
    \caption{Coronal view.}
    \label{fig:sim_pm_coronal}
\end{subfigure}\hfill
\begin{subfigure}[b]{0.32\columnwidth}
    \centering
    \includegraphics[width=\linewidth]{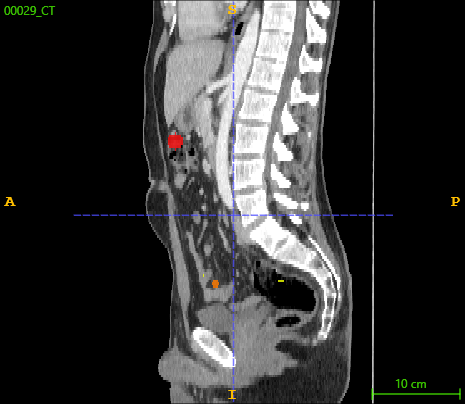}
    \caption{Sagittal view.}
    \label{fig:sim_pm_sagittal}
\end{subfigure}

\caption{Example of simulated peritoneal metastases. Colored overlays represent nodules of different PCI lesion-size classes (LS1–LS3) within the CT scan.}
\label{fig:sim_pm_example}
\end{figure}

\noindent\textbf{Clinical Variability Simulation:}
Geometric segmentation metrics do not directly capture the clinical relevance of segmentation errors, as treatment planning relies on the PCI rather than precise segmentation boundaries. To investigate how improvements in typically used segmentation metrics translate into better clinical decision making, a probabilistic simulation of PM was implemented on majority-vote label maps by sampling lesion voxels within rPCI regions and then querying the region ID assigned by each observer/model at those voxels to quantify region-assignment disagreement.
These label maps were constructed by determining, for each voxel, which of the 13 region labels was assigned by at least two out of three annotators. For each trial, the number of nodules was drawn from a Poisson distribution with a mean of 15, and clipped to the range~1--50. This range represents a moderate disease burden while avoiding implausible extremes~\cite{Bhatt2024}. Simulated nodules were assigned to rPCI regions using Dirichlet-weighted regional probabilities based on reported distribution patterns from literature~\cite{Jacquet1996ClinicalCarcinomatosis,Wong2022}. Within each selected region, a voxel was sampled from the 3D label map as the nodule location, and a lesion-size class~(LS0--LS3) was assigned uniformly.
Figure~\ref{fig:sim_pm_example} shows an example case with simulated PM nodules in the abdominal region of a patient.
For each simulation, the rPCI was calculated by selecting the largest lesion in each region, assigning a lesion-size class, and summing across all regions. This total was used as the simulated reference (ground-truth) rPCI. At each simulated nodule location, the region label assigned by each observer and by the nnU-Net segmentation was recorded, and a predicted rPCI was computed from these labels. Region-level accuracy was assessed by comparing predicted and ground-truth region labels at each nodule location. At the patient level, an rPCI of~20 was used as a clinical threshold to classify cases as high or low rPCI, and sensitivity and specificity were calculated at this cutoff. The distributions of rPCI differences between simulated reference and predicted rPCI, along with these metrics, were analyzed around the rPCI cutoff of~20.

\section{Results}
\label{sec:res}
\begin{figure}[t]
\centering
\begin{subfigure}[c]{0.32\columnwidth}
    \centering
    \includegraphics[width=\linewidth,angle=180]{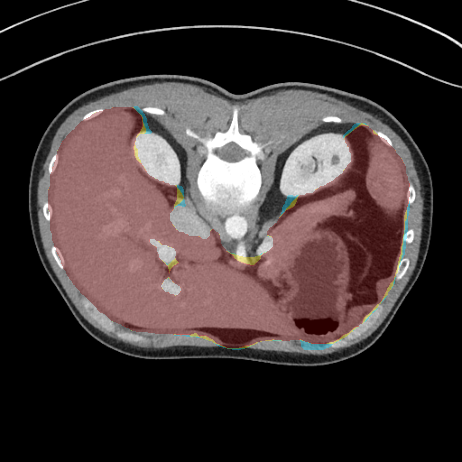}
    \caption{Axial view.}
    \label{fig:qualitative_axial}
\end{subfigure}\hfill
\begin{subfigure}[c]{0.32\columnwidth}
    \centering
    \includegraphics[width=\linewidth]{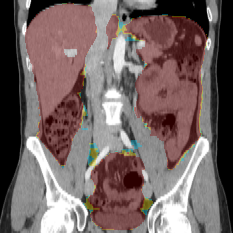}
    \caption{Coronal view.}
    \label{fig:qualitative_coronal}
\end{subfigure}\hfill
\begin{subfigure}[c]{0.32\columnwidth}
    \centering
    \includegraphics[width=\linewidth]{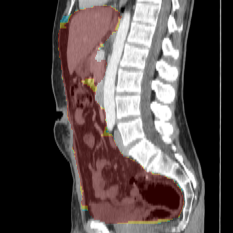}
    \caption{Sagittal view.}
    \label{fig:qualitative_sagittal}
\end{subfigure}
\caption{Qualitative inter-observer agreement heatmaps and contour overlays.}
\label{fig:qualitative_maps}
\end{figure}

\noindent\textbf{Inter-Observer Variability:} Observers demonstrated high overall agreement across the 13~rPCI regions. Across both cohorts, the mean inter-observer Dice was $0.87 \pm 0.04$, HD95 was $8.8 \pm 3.0$\,mm, and ASD was $2.4 \pm 1.1$\,mm. Variability was concentrated in the small-bowel regions~(9--12): Dice decreased from $0.88 \pm 0.05$ (regions~0--8) to $0.86 \pm 0.01$ (regions~9--12), while HD95 increased from $7.5 \pm 2.7$ to $11.6 \pm 1.4$\,mm and ASD from $1.7 \pm 0.6$ to $3.8 \pm 0.6$\,mm. Qualitative disagreement maps confirmed that differences were spatially localized around the small-bowel, while upper abdominal regions exhibited nearly identical contours (Fig.~\ref{fig:qualitative_maps}).

\begin{figure}[t]
    \centering    
    \includegraphics[width=0.91\columnwidth,trim=16mm 21mm 16mm 22mm,clip]{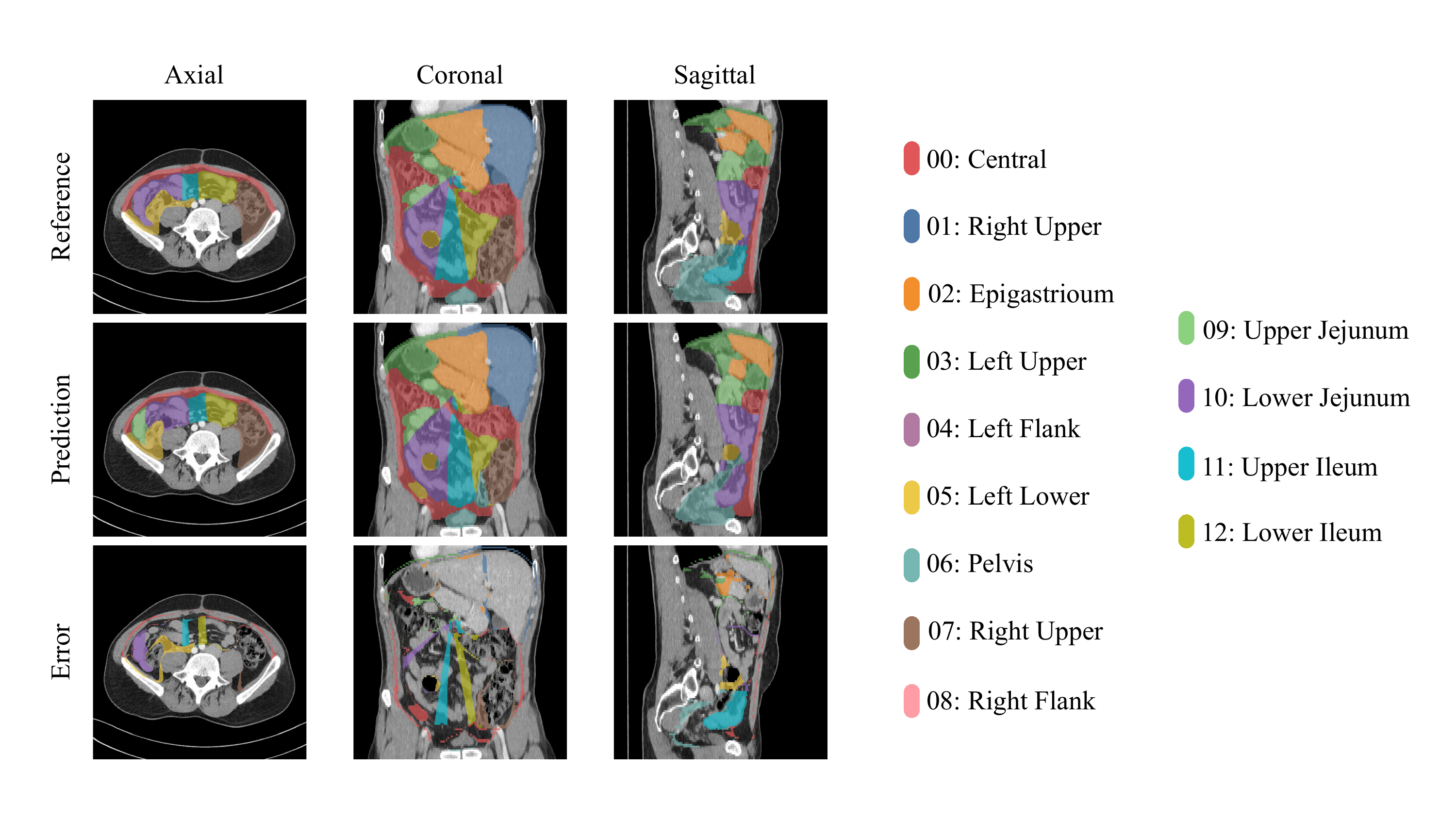}
    \caption{Qualitative CT overlays comparing majority-vote reference segmentation, model prediction, and segmentation error in axial, coronal, and sagittal slices.}
    \label{fig:modelhuman_disagreement}
\end{figure}

\noindent\textbf{Human--model variability:} Model--observer agreement was lower across all metrics: mean Dice $0.82 \pm 0.06$, HD95 $20.8 \pm 17.0$\,mm, and ASD $4.4 \pm 2.4$\,mm. The model aligned more closely with observers in regions 0--8 (Dice $0.84 \pm 0.06$, HD95 $15.8 \pm 12.5$\,mm, ASD $3.3 \pm 1.5$\,mm) than in regions 9--12 (Dice $0.77 \pm 0.04$, HD95 $31.9 \pm 22.4$\,mm, ASD $6.9 \pm 2.3$\,mm). As shown in Table~\ref{tab:regionlevel}, regions 4, 8, 11, and 12 show the largest deviations, with reduced Dice and increased HD95/ASD. The latter suggests localized boundary errors rather than uniform over- or under-segmentation. Figure~\ref{fig:modelhuman_disagreement} illustrates this pattern, with most model errors occurring in the small-bowel area and at boundaries between rPCI regions.
To evaluate consistency across scans, paired differences (M$-$H) were tested using a two-sided Wilcoxon signed-rank test ($N=10$), supporting differences for Dice ($p=0.0273$) and HD95 ($p=0.0195$), with weaker evidence for ASD ($p=0.1289$).

\newcommand{\pms}[2]{#1\!\pm\!#2}
\begin{table*}[b]
\centering
\caption{Region-level inter-observer (H) vs.\ model--observer (M) agreement (mean $\pm$ SD); HD95/ASD in mm; $\Delta$ = M$-$H.}
\label{tab:regionlevel}
\begingroup
\fontsize{8}{9}\selectfont
\setlength{\tabcolsep}{1.9pt}
\renewcommand{\arraystretch}{0.90}
\begin{tabular}{@{}c|ccccccccc@{}}
\toprule
Region
& Dice$_H$
& Dice$_M$
& $\Delta$Dice
& HD95$_H$
& HD95$_M$
& $\Delta$HD95
& ASD$_H$
& ASD$_M$
& $\Delta$ASD \\
\midrule
0  & $\pms{0.86}{0.03}$ & $\pms{0.81}{0.04}$ & $-0.05$ & $\pms{6.3}{1.4}$  & $\pms{11.7}{4.4}$ & $5.4$  & $\pms{1.6}{0.4}$ & $\pms{2.8}{0.7}$  & $1.2$ \\
1  & $\pms{0.96}{0.01}$ & $\pms{0.94}{0.01}$ & $-0.02$ & $\pms{4.8}{1.2}$  & $\pms{6.2}{2.2}$  & $1.5$  & $\pms{1.0}{0.3}$ & $\pms{1.9}{0.6}$  & $0.8$ \\
2  & $\pms{0.91}{0.01}$ & $\pms{0.88}{0.02}$ & $-0.03$ & $\pms{6.2}{1.4}$  & $\pms{8.8}{3.0}$  & $2.6$  & $\pms{1.4}{0.3}$ & $\pms{2.4}{0.7}$  & $1.0$ \\
3  & $\pms{0.94}{0.01}$ & $\pms{0.91}{0.02}$ & $-0.03$ & $\pms{4.8}{0.8}$  & $\pms{7.2}{2.5}$  & $2.3$  & $\pms{1.1}{0.2}$ & $\pms{2.0}{0.6}$  & $0.8$ \\
4  & $\pms{0.85}{0.05}$ & $\pms{0.78}{0.07}$ & $-0.07$ & $\pms{10.4}{3.1}$ & $\pms{30.8}{10.6}$& $20.4$ & $\pms{2.2}{0.5}$ & $\pms{6.1}{2.1}$  & $3.9$ \\
5  & $\pms{0.87}{0.04}$ & $\pms{0.81}{0.05}$ & $-0.07$ & $\pms{4.6}{0.9}$  & $\pms{7.7}{2.4}$  & $3.1$  & $\pms{1.3}{0.2}$ & $\pms{2.0}{0.6}$  & $0.6$ \\
6  & $\pms{0.89}{0.02}$ & $\pms{0.89}{0.03}$ & $-0.01$ & $\pms{10.0}{3.2}$ & $\pms{13.6}{2.6}$ & $3.6$  & $\pms{2.5}{0.8}$ & $\pms{3.6}{0.6}$  & $1.1$ \\
7  & $\pms{0.84}{0.05}$ & $\pms{0.81}{0.05}$ & $-0.04$ & $\pms{10.1}{4.9}$ & $\pms{14.0}{5.7}$ & $3.9$  & $\pms{2.1}{0.8}$ & $\pms{3.9}{2.3}$  & $1.8$ \\
8  & $\pms{0.80}{0.08}$ & $\pms{0.77}{0.06}$ & $-0.03$ & $\pms{10.6}{6.1}$ & $\pms{42.6}{36.6}$& $32.0$ & $\pms{2.5}{1.1}$ & $\pms{5.4}{2.4}$  & $2.9$ \\
9  & $\pms{0.87}{0.07}$ & $\pms{0.82}{0.06}$ & $-0.05$ & $\pms{10.7}{6.2}$ & $\pms{26.4}{22.2}$& $15.7$ & $\pms{3.5}{1.2}$ & $\pms{5.4}{2.9}$  & $1.9$ \\
10 & $\pms{0.86}{0.06}$ & $\pms{0.78}{0.08}$ & $-0.08$ & $\pms{11.9}{6.5}$ & $\pms{18.8}{6.8}$ & $6.9$  & $\pms{3.5}{1.4}$ & $\pms{5.6}{1.7}$  & $2.1$ \\
11 & $\pms{0.84}{0.05}$ & $\pms{0.73}{0.12}$ & $-0.11$ & $\pms{10.3}{3.5}$ & $\pms{17.5}{4.6}$ & $7.3$  & $\pms{3.3}{1.2}$ & $\pms{6.2}{2.2}$  & $2.8$ \\
12 & $\pms{0.85}{0.04}$ & $\pms{0.75}{0.08}$ & $-0.10$ & $\pms{13.4}{3.3}$ & $\pms{65.0}{42.6}$& $51.6$ & $\pms{4.7}{1.6}$ & $\pms{10.2}{4.0}$ & $5.4$ \\
\bottomrule
\end{tabular}
\endgroup
\end{table*}

\begin{figure}[t]
\centering
\begin{subfigure}[t]{0.48\columnwidth}
    \centering
    \includegraphics[width=\linewidth,trim=0.5mm 2.5mm 2.5mm 2mm,clip]{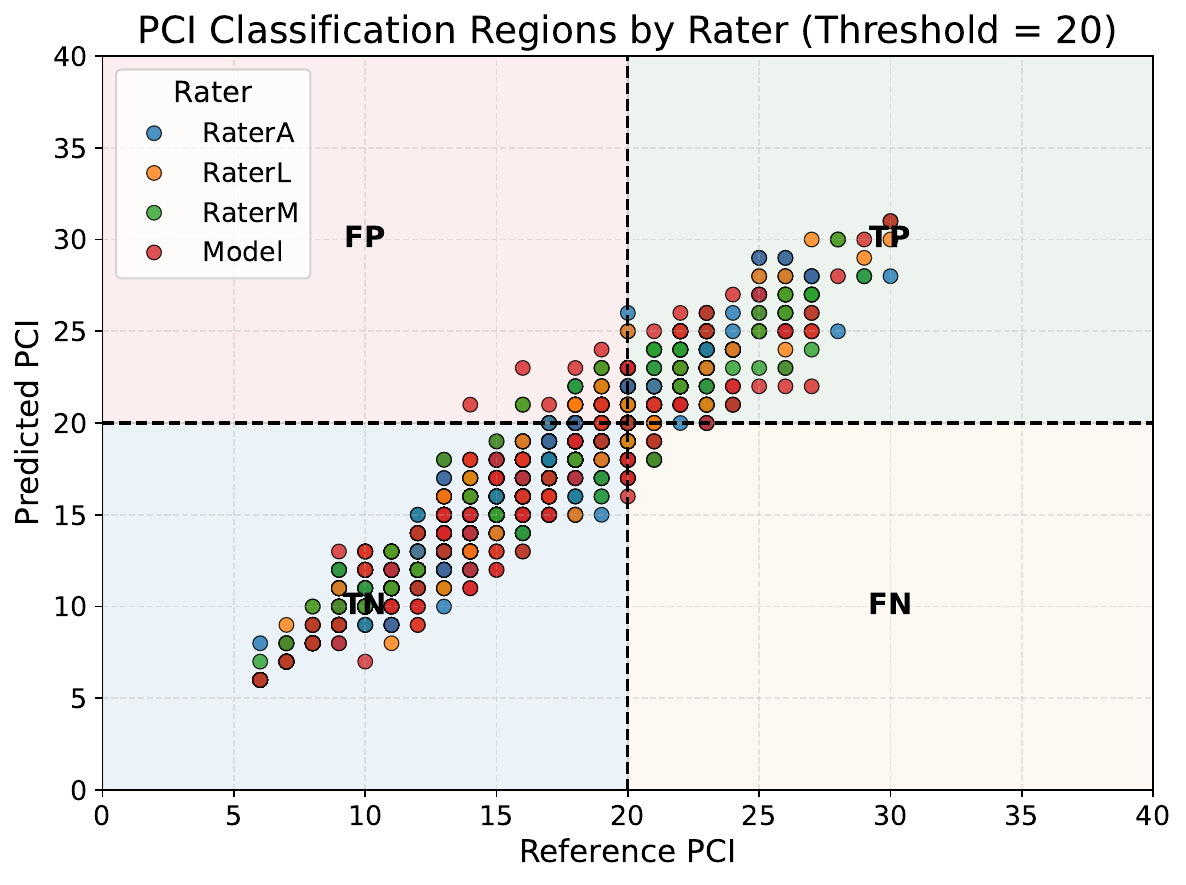}
    \label{fig:pci_cutoff_scatter_A}
\end{subfigure}\hfill
\begin{subfigure}[t]{0.47\columnwidth}
    \centering
    \includegraphics[width=\linewidth,trim=0.5mm 2.5mm 2.5mm 2mm,clip]{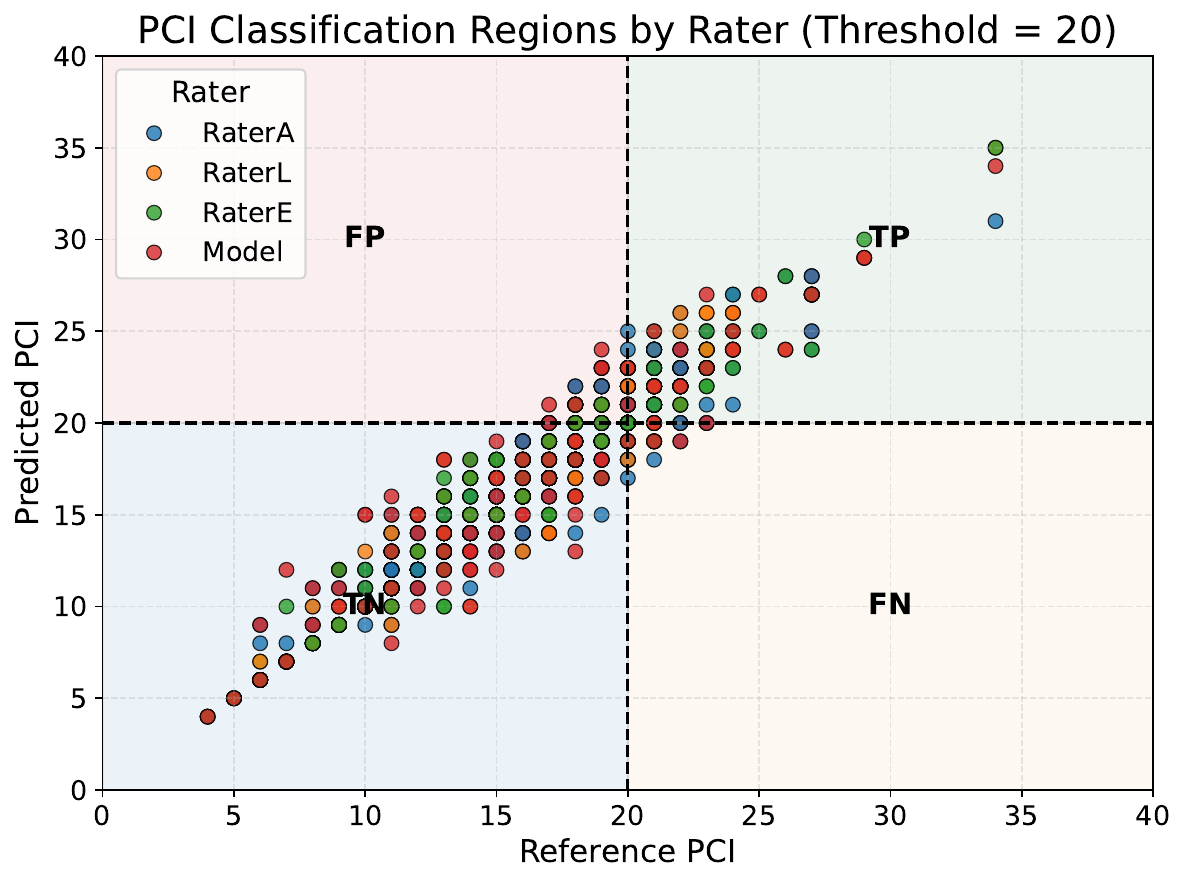}
    \label{fig:pci_cutoff_scatter_B}
\end{subfigure}
\caption{Cohorts A \& B: Predicted vs.\ reference PCI at a cutoff of 20, categorized into TN, TP, FP, and FN.}
\label{fig:pci_cutoff_scatter}
\end{figure}

\noindent\textbf{Clinical variability:} Region-level accuracy was high in both cohorts, ranging from $0.90$ to $0.92$ for human observers and $0.83$--$0.85$ for the model, and the mean $\Delta$~rPCI remained below one point for all observers and the model in both cohorts (Table~\ref{tab:pci_summary}), indicating that segmentation differences generally resulted in minor changes in rPCI. At the PCI~20 threshold, sensitivity ranged from $0.87$--$0.98$ and specificity from $0.92$--$0.96$ across observers and the model in both cohorts (Table~\ref{tab:pci_summary}), with misclassifications concentrated near the decision boundary (Fig.~\ref{fig:pci_cutoff_scatter}), where small changes in rPCI can alter treatment eligibility.

\begin{table}[t]
\centering
\caption{Summary of simulated clinical variability: region-level accuracy and rPCI differences (mean $\pm$ SD) and classification performance at PCI~20.}
\label{tab:pci_summary}
\begin{subtable}[t]{0.48\columnwidth}
\centering
\caption{Cohort A}
\label{tab:pci_summary_A}
\resizebox{\columnwidth}{!}{%
\begin{tabular}{l|cccc}
\toprule
Observer & Accuracy & $\Delta$rPCI & Sens. & Spec. \\
\midrule
RaterA & $0.92 \pm 0.08$ & $0.35 \pm 1.24$ & $0.93$ & $0.96$ \\
RaterL & $0.92 \pm 0.07$ & $0.45 \pm 1.15$ & $0.96$ & $0.94$ \\
RaterM & $0.90 \pm 0.09$ & $0.32 \pm 1.18$ & $0.89$ & $0.94$ \\
Model  & $0.85 \pm 0.10$ & $0.34 \pm 1.55$ & $0.87$ & $0.93$ \\
\bottomrule
\end{tabular}%
}
\end{subtable}\hfill
\begin{subtable}[t]{0.48\columnwidth}
\centering
\caption{Cohort B}
\label{tab:pci_summary_B}
\resizebox{\columnwidth}{!}{%
\begin{tabular}{l|cccc}
\toprule
Observer & Accuracy & $\Delta$rPCI & Sens. & Spec. \\
\midrule
RaterA & $0.88 \pm 0.09$ & $0.45 \pm 1.38$ & $0.92$ & $0.94$ \\
RaterL & $0.91 \pm 0.08$ & $0.46 \pm 1.18$ & $0.93$ & $0.96$ \\
RaterE & $0.87 \pm 0.12$ & $0.49 \pm 1.14$ & $0.98$ & $0.95$ \\
Model  & $0.83 \pm 0.10$ & $0.59 \pm 1.58$ & $0.92$ & $0.92$ \\
\bottomrule
\end{tabular}%
}
\end{subtable}
\end{table}

\section{Discussion}
\label{sec:disc}

\noindent\textbf{Contribution and decision-level framing:}
This study does not propose a new segmentation architecture or training strategy. Instead, it addresses a translation gap: standard segmentation metrics quantify geometric agreement but do not indicate whether disagreements would change rPCI scores or alter treatment eligibility. The contribution is a reproducible evaluation framework that propagates segmentation variability into PCI variability and tests impact at a clinically used threshold. This provides decision-level evidence to support deployment decisions and to define when expert review is most critical, particularly for borderline cases.

\noindent\textbf{Variability patterns across regions and metrics:}
Strong inter-observer agreement was found in most rPCI regions, with higher variation in the small-bowel segments, consistent with the challenges of abdominal CT interpretation. The model showed a similar spatial error pattern but with larger deviations in specific regions, suggesting that anatomical ambiguity remains a shared bottleneck for manual and automated segmentation. The model’s regional error pattern only partly aligns with the Delphi consensus on rPCI definitions~\cite{TopsWelten2025}: both highlight small-bowel challenges, but the model also struggled in regions not flagged by the Delphi process (e.g.\ regions 4 and 8) while human observers remained consistent. This indicates that model sensitivity to local contrast, bowel motion, and anatomical variation contributes beyond definitional ambiguity alone. A divergence between overlap and surface-distance metrics was also observed in these regions: ASD increased moderately from inter-observer to model--observer comparisons, whereas HD95 increased substantially more. This pattern suggests small, isolated model misclassifications far from the true boundary, which inflate HD95 but have limited impact on PCI scoring.

\noindent\textbf{Impact on rPCI scoring and the clinical threshold:}
The resulting rPCI differences were small (typically $<1$ point on average), indicating that segmentation variability rarely shifts cases far from the decision boundary. It is important to separate this region-assignment variability from the broader radiological-to-surgical PCI gap, where missed lesions and under-staging may dominate. Threshold analysis showed that misclassifications clustered near PCI~20, where small rPCI shifts can change treatment eligibility. The model’s high specificity but lower sensitivity suggests a conservative bias around the threshold, with borderline and high-rPCI cases more often predicted just below 20. In practice, region segmentations can support a structured, region-by-region reading workflow, acting as a checklist to help assign findings to rPCI regions and potentially reduce oversight and reader workload.

\noindent\textbf{Limitations:}
Although the cohorts are small relative to population-level studies, the study yields 130 region-level comparisons per observer pair (13 regions $\times$ 10 scans), and the simulation generates thousands of rPCI trials per scan, providing dense sampling of the variability landscape. The objective is not to estimate population-level rPCI performance, but to characterize decision robustness under controlled conditions; consistency across two independent cohorts strengthens the conclusions beyond what a single small dataset would allow. The majority-vote reference is derived from three annotators within each cohort, so each rater is compared against a consensus they helped create, which may modestly inflate inter-observer agreement. The nnU-Net model does not contribute to the majority vote and is therefore disadvantaged, so model results should be interpreted in that context.
The simulation links rPCI boundary disagreement to PCI variability, but relies on simplifying assumptions. Lesions are placed using fixed, region-specific probabilities and scored by point-wise label lookup, whereas real deposits can cluster near region borders, which may bias border-related assignment errors. Lesions are localized by a single voxel within the majority-vote masks, avoiding ambiguity from lesion extent; in practice, extended lesions may span multiple regions and require an assignment rule. Lesion sizes are sampled uniformly, although smaller deposits are more common clinically and could influence threshold behavior. Finally, acquisition factors and reader-dependent effects that drive missed lesions and under-staging are not modeled and likely explain much of the radiological-to-surgical PCI gap. The reported PCI variability should therefore be interpreted as a lower bound attributable to region segmentation variability alone. Prospective validation against surgical PCI in larger multi-center cohorts is needed to capture the combined effects of detectability, missed lesions, and region-assignment variability.

\section{Conclusion}
\label{sec:con}
This study provides the first quantification of inter-observer variability in rPCI region segmentation and introduces a simulation framework that propagates this variability into PCI scores to evaluate its impact at a clinical decision threshold. Model-induced variability was generally within the human inter-observer range in well-defined anatomical regions, but exceeded it in several regions. Despite these geometric differences, simulated PCI scores deviated by less than one point on average, and most predictions were classified on the correct side of the PCI~20 threshold, with misclassifications concentrated near borderline cases. Since the simulation isolates region-assignment variability and does not account for missed lesions or acquisition-related factors, these deviations represent a lower bound on total rPCI uncertainty. Together, these results support rPCI as a viable tool for non-invasive PCI estimation, while cases near clinically relevant thresholds still require careful expert review. Future work should validate model-assisted scoring against surgical PCI in larger, multi-center cohorts, and complement region segmentation with lesion detection models targeting small or low-contrast tumor deposits.

\bibliographystyle{splncs04}
\bibliography{bibliography}

\end{document}